\documentclass[prb,twocolumn,showpacs]{revtex4}
\usepackage{epsfig}
\usepackage{graphicx}
\usepackage{soul}
\usepackage{color}
\usepackage{wrapfig}
\usepackage{amsmath}
\begin{document}
%%%%%%%%%%%%%%%%%%%%%%%%%%%%%%%%%%%%%%%%%%%%%%%%%%%%%%%%%%%%%%%%%%%%%%%%%%%%%%%%
\title{Topological split-ring resonator based metamaterials with $\cal PT$
symmetry relying on gain and loss}
%%%%%%%%%%%%%%%%%%%%%%%%%%%%%%%%%%%%%%%%%%%%%%%%%%%%%%%%%%%%%%%%%%%%%%%%%%%%%%%%
\author{N. Lazarides$^{1,2}$, G. P. Tsironis$^{1,2}$}
\affiliation{
$^{1}$Institute of Theoretical and Computational Physics $\&$ 
      Department of Physics, University of Crete, Herakleio, Greece \\
$^{2}$National University of Science and Technology ``MISiS'', Leninsky Prospekt 
      4, Moscow, 119049, Russia
}
\date{\today}
%%%%%%%%%%%%%%%%%%%%%%%%%%%%%%%%%%%%%%%%%%%%%%%%%%%%%%%%%%%%%%%%%%%%%%%%%%%%%%%%
\begin{abstract}
A one-dimensional metamaterial with parity-time (${\cal PT}$) symmetry that 
relies on balanced gain and loss is introduced, comprising of magnetically 
coupled split-ring resonators (SRRs). A particular topology that combines a 
non-trivial (topological) dimer configuration with a trivial (non-topological) 
dimer configuration which are separated by a central SRR with neither gain or 
loss, is investigated. By focusing on the dynamical aspects of such a 
topological ${\cal PT}$ metamaterial (PTMM), the existence of {\em topologically
protected interface states} which are localized at the central SRR is 
demonstrated numerically. The solution of the corresponding {\em quadratic 
eigenvalue problem} reveals that the protected state is actually a robust 
eigenmode of the topological PTMM, whose eigenvalue is isolated in the middle of 
the gap (mid-gap state) of the two-band frequency spectrum. Direct numerical 
simulations have been further used to determine the robustness and dynamic 
stability of these states in the parameter space of the {\em dimerization 
strength} and the {\em gain-loss coefficient}.
\end{abstract}
%%%%%%%%%%%%%%%%%%%%%%%%%%%%%%%%%%%%%%%%%%%%%%%%%%%%%%%%%%%%%%%%%%%%%%%%%%%%%%%%
\pacs{63.20.Pw, 11.30.Er, 41.20.-q, 78.67.Pt}
%%%%%%%%%%%%%%%%%%%%%%%%%%%%%%%%%%%%%%%%%%%%%%%%%%%%%%%%%%%%%%%%%%%%%%%%%%%%%%%%
\maketitle
%%%%%%%%%%%%%%%%%%%%%%%%%%%%%%%%%%%%%%%%%%%%%%%%%%%%%%%%%%%%%%%%%%%%%%%%%%%%%%%%
\section{Introduction}
%%%%%%%%%%%%%%%%%%%%%%%%%%%%%%%%%%%%%%%%%%%%%%%%%%%%%%%%%%%%%%%%%%%%%%%%%%%%%%%%
The application of topology, the mathematical theory of conserved properties 
under continuous deformations, is creating new opportunities for several fields 
including photonics \cite{LLu2014,Noh2018,Ozawa2019}, acoustics 
\cite{YGPeng2018,Apigo2019,Ma2019}, and mechanics 
\cite{Kane2013,Rocklin2016,Prodan2017,Ma2019}. This field was inspired by the 
discovery of {\em topological insulators} 
\cite{Moore2010,Hasan2010,Rachel2018,Ma2019}, in which interfacial electrons 
transport without dissipation, even in the presence of impurities. Similarly, 
using carefully designed topologies allows the creation of interfaces that 
support new protected states in photonics \cite{Schomerus2013,Weimann2017} and 
elastic photonic crystals \cite{JYin2018}.

Recently, the topological properties of photonic $\cal PT-$symmetric crystals 
\cite{Weimann2017} and other non-Hermitian systems 
\cite{Esaki2011,Lee2016,Lang2018,Takata2018,KKawabata2019,XWLuo2019,Kottos2020,Slager2020} 
have attracted a lot of interest. Moreover, the topological insulating properties 
\cite{TCao2013,Barlas2020} and the topological transitions 
\cite{Krishnamoorthy2012,CLiu2017} of various metamaterials have been 
investigated. Further, topological mechanical metamaterials \cite{Meeussen2020} 
and topological metamaterials based on polariton rings \cite{Kozin2018} were 
introduced. These and similar concepts were recently transferred into electronic 
and electrical circuit devices \cite{Kotwal2019,Voss2020} and also in split-ring 
resonator (SRR) chains \cite{JJiang2018,JJiang2020}, were robust topologically 
protected states were detected.

\begin{figure}[h!]
\includegraphics[angle=0,width=1.0 \linewidth]{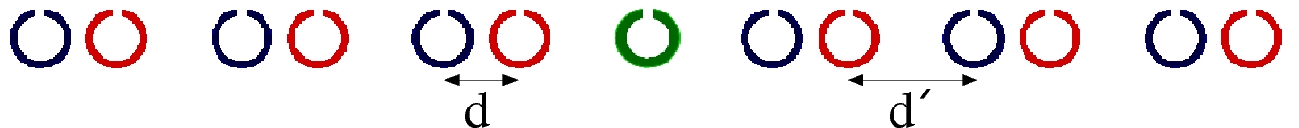}
\caption{(color online) 
Schematic of the topological $\cal PT-$ metamaterial. The blue, red, and green 
split-ring resonator(s) have loss, gain, and neither gain nor loss, 
respectively. SRRs separated by center-to-center distance $d$ ($d'$) are coupled 
magnetically and electrically through intra-dimer (inter-dimer) coefficients 
$\lambda_M$ and $\lambda_E$ ($\lambda_M'$ and $\lambda_E'$), respectively.
}
\label{fig01}
\end{figure}

Here, a one-dimensional (1D) parity-time (${\cal PT}$) symmetric metamaterial 
(PTMM) comprising SRRs which are magnetically coupled through dipole-dipole 
interactions due to their mutual inductance 
\cite{Lazarides2006,Lazarides2013a,Tsironis2014,Lazarides2019}, is considered. 
The exact configuration (Fig. \ref{fig01}) combines a topological with a 
non-topological dimer chain of SRRs, both having alternating gain and loss. The 
chains are separated by a central SRR with neither gain or loss 
\cite{Weimann2017}. Note that non-topological PTMMs support localized states 
either due to nonlinearity \cite{Lazarides2013a,Tsironis2014} or due to a flat 
band in their spectrum \cite{Lazarides2019}. Symmetry-protected localized states 
whose localization lengths are robust against gain and loss have been 
investigated in $\cal PT-$symmetric ladder lattices \cite{Ryu2019}, passive 
$\cal PT-$symmetric microwave resonator Su-Schrieffer-Heeger (SSH) chains 
\cite{Poli2015}, complex photonic lattices with spatially distributed gain and 
loss \cite{Schomerus2013}, SSH chains with a pair of $\cal PT-$symmetric 
defects \cite{Jin2017}, non-Hermitian trimerized optical lattices 
\cite{Jin2017b}, and complex SSH lattice models containing gain and loss on 
alternating sites \cite{Lang2018}.

The present work focuses on the dynamical aspects of the topological PTMM
in Fig. \ref{fig01}, and in particular on the topological protection of 
localized interface states in the parameter space of the gain-loss coefficient
$\gamma$ and the dimerization strength 
$\delta \lambda_M = |\lambda_M -\lambda_M'|$, with $\lambda_M$ ($\lambda_M'$) 
being the intra-dimer (inter-dimer) magnetic coupling strength between 
neighboring SRRs separated by center-to-center distance $d$ ($d'$). Although the 
coupling between SRRs is generally both electric and magnetic, the former can be 
much weaker than the latter under certain conditions. Here, the electric 
coupling is neglected for simplicity; however, qualitatively similar results are 
obtained if both couplings are taken into account. 
%%%%%%%%%%%%%%%%%%%%%%%%%%%%%%%%%%%%%%%%%%%%%%%%%%%%%%%%%%%%%%%%%%%%%%%%%%%%%%%%
%%%%%%%%%%%%%%%%%%%%%%%%%%%%%%%%%%%%%%%%%%%%%%%%%%%%%%%%%%%%%%%%%%%%%%%%%%%%%%%%

In the next Section, the dynamic equations for the topological PTMM are 
presented in matrix form, along with a general description of the equivalent 
circuit model. In Section III, those equations are simulated with single-site 
initial excitations, to demonstrate the emergence of topologically protected 
localized interface states. The corresponding quadratic eigenvalue problem is 
solved in Section IV, revealing that the protected states are robust eigenmodes 
with an isolated midgap eigenvalue. Conclusions are given in Section V.

%%%%%%%%%%%%%%%%%%%%%%%%%%%%%%%%%%%%%%%%%%%%%%%%%%%%%%%%%%%%%%%%%%%%%%%%%%%%%%%%
\section{Model Dynamical Equations}
%%%%%%%%%%%%%%%%%%%%%%%%%%%%%%%%%%%%%%%%%%%%%%%%%%%%%%%%%%%%%%%%%%%%%%%%%%%%%%%%
Consider the topological PTMM of Fig. \ref{fig01}, in which the SRRs have either 
loss (blue), or gain (red), or neither of those (green). Their total number is 
$N=5+4 m$, where $m$ positive integer or zero. In the equivalent circuit model 
picture, the SRRs are regarded as resistive-inductive-capacitive (RLC) circuits, 
featuring  resistance $R$, inductance $L$, and capacitance $C$. Note that the 
resistance is positive, $+R$ (negative, $-R$) for the SRRs with loss (gain), 
while it is zero for the central SRR. The SRRs are considered to be coupled 
magnetically to their nearest neighbors through their mutual inductance $M$ and 
$M'$ for separating center-to-center distance $d$ and $d'$, respectively 
($d < d'$). The intra-dimer (inter-dimer) coupling coefficients, defined as 
$\lambda_M =M/L$ ($\lambda_M' =M'/L$), usually assume rather low values that 
justify the nearest-neighbor coupling approximation. An initial excitation 
induces currents in the SRRs which either decay or amplify depending on them 
having respectively loss or gain. However, the balance of gain and loss can lead 
the system in a steady state in which periodic solutions exist as long as the 
gain-loss coefficient $\gamma$ remains below its critical value $\gamma_c$ that 
separates the exact from the broken $\cal PT$ phase.

The dynamics of the topological PTMM is determined by the evolution of the 
charges $Q_n$ ($n=1,2,3, ... ,N$) accumulated at the banks of the capacitors $C$ 
of the SRRs. In order to exemplify the model, let us consider the SRR at 
$n =n_e -1$ (i.e., the left nearest neighbor of the central SRR in Fig. 
\ref{fig01}), whose dynamic equation can be derived from Kirchhoff's voltage low 
and the relation $I_{n_e -1} =d Q_{n_e -1} /dt$, as  
\begin{eqnarray}
\nonumber
\frac{d^2}{dt^2} \left( M Q_{n_e -2} +L Q_{n_e -1} +M' Q_{n_e} \right)
-R \frac{d Q_{n_e -1}}{dt} 
\\ \nonumber 
+\frac{Q_{n_e -1}}{C} =0,
\end{eqnarray}
where $I_{n_e -1}$ is the current in the SRR at $n=n_e -1$. That equation can be 
casted in the dimensionless form 
\begin{eqnarray}
\nonumber
\frac{d^2}{d\tau^2} 
\left( \lambda_M q_{n_e -2} +q_{n_e -1} +\lambda_M' q_{n_e} \right)
-\gamma \frac{d q_{n_e -1}}{d\tau} 
\\ \nonumber
+q_{n_e -1} =0,
\end{eqnarray}
using the relations $\tau =\omega_{LC} t$ and $q_{n_e -1} =Q_{n_e -1}/(C U_0)$, 
where $U_0$ is a characteristic voltage across the slits of the SRRs, 
$\gamma =R \sqrt{C/L}$, and $\omega_{LC} =1/\sqrt{L C}$ is the 
inductive-capacitive ($L\,C$) resonance frequency of the SRRs.

By carefully considering the coupling coefficients and the sign in front of 
$\gamma$ or the absence of $\gamma$ (for the central SRR), similar equations can 
be derived for all the SRRs of the topological PTMM. This procedure results in a 
modification of PTMM models employed elsewhere 
\cite{Lazarides2013a,Tsironis2014,Lazarides2019}, that accounts for the 
configuration considered here. The equations of that model can be written in 
compact, matrix form as
\begin{equation}
\label{eq01}
   \hat{\bf{\Lambda}}_M \ddot{\bf q} 
  +\hat{\bf{\Gamma}} \dot{\bf {q}} +{\bf q} ={\bf 0},
\end{equation}
where ${\bf q} =[q_1 ~q_2  ~...  ~q_N]^T$ is an $N-$dimensional vector, the 
overdots denote differentiation over the normalized temporal variable $\tau$, 
and the $N\times N$ matrices $\hat{\bf{\Lambda}}_M$ and $\hat{\bf{\Gamma}}$ 
are given by
\begin{eqnarray} 
\label{eq09}
\hat{\bf \Lambda}_M=
  \begin{bmatrix}
    1                    & \lambda_M            & 0                    & 0                    & \reflectbox{$\ddots$}& 0                    & 0                    & 0                    & 0 \\
    \lambda_M            & 1                    & \lambda_M'           & \reflectbox{$\ddots$}& 0                    & 0                    & 0                    & 0                    & 0 \\
    0                    & \lambda_M'           & \reflectbox{$\ddots$}& \lambda_M            & 0                    & 0                    & 0                    & 0                    & 0 \\
    0                    & \reflectbox{$\ddots$}& \lambda_M            & 1                    & \lambda_M'           & 0                    & 0                    & 0                    & 0 \\ 
    \reflectbox{$\ddots$}& 0                    & 0                    & \lambda_M'           & 1                    & \lambda_M'           & 0                    & 0                    & \reflectbox{$\ddots$} \\
    0                    & 0                    & 0                    & 0                    & \lambda_M'           & 1                    & \lambda_M            & \reflectbox{$\ddots$}& 0 \\
    0                    & 0                    & 0                    & 0                    & 0                    & \lambda_M            & \reflectbox{$\ddots$}& \lambda_M'           & 0 \\
    0                    & 0                    & 0                    & 0                    & 0                    & \reflectbox{$\ddots$}& \lambda_M'           & 1                    & \lambda_M \\
    0                    & 0                    & 0                    & 0                    & \reflectbox{$\ddots$}& 0                    & 0                    & \lambda_M            & 1
  \end{bmatrix} , 
\end{eqnarray} 
and
\begin{eqnarray} 
\label{eq10}
\hat{\bf \Gamma}=
  \begin{bmatrix}
    +\gamma              & 0            & 0                    & 0                     & \reflectbox{$\ddots$}& 0                    & 0                    & 0                    & 0 \\
    0                    & -\gamma              & 0                    & \reflectbox{$\ddots$} & 0                    & 0                    & 0                    & 0                    & 0 \\
    0                    & 0                    & \reflectbox{$\ddots$}& 0                     & 0                    & 0                    & 0                    & 0                    & 0 \\
    0                    & \reflectbox{$\ddots$}& 0                    & -\gamma               & 0                    & 0                    & 0                    & 0                    & 0 \\ 
    \reflectbox{$\ddots$}& 0                    & 0                    & 0                     & 0                    & 0                    & 0                    & 0                    & \reflectbox{$\ddots$} \\
    0                    & 0                    & 0                    & 0                     & 0                    & +\gamma              & 0                    & \reflectbox{$\ddots$}& 0 \\
    0                    & 0                    & 0                    & 0                     & 0                    & 0                    & \reflectbox{$\ddots$}& 0                    & 0 \\
    0                    & 0                    & 0                    & 0                     & 0                    & \reflectbox{$\ddots$}& 0                    & +\gamma              & 0 \\
    0                    & 0                    & 0                    & 0                     & \reflectbox{$\ddots$}& 0                    & 0                    & 0                    & -\gamma
  \end{bmatrix}. 
\end{eqnarray} 
Note that the two SRRs at each end of the chain interact through intra-dimer 
coupling $\lambda_M$, excluding thus the formation of topologically protected 
{\em edge states}. The coefficients $\lambda_M$ and $\lambda_M'$ can be either 
estimated by simple means \cite{Lazarides2006}, or calculated accuratelly using 
commercially available software packages \cite{Rosanov2011}.

%%%%%%%%%%%%%%%%%%%%%%%%%%%%%%%%%%%%%%%%%%%%%%%%%%%%%%%%%%%%%%%%%%%%%%%%%%%%%%%%
\section{Numerical simulations}
%%%%%%%%%%%%%%%%%%%%%%%%%%%%%%%%%%%%%%%%%%%%%%%%%%%%%%%%%%%%%%%%%%%%%%%%%%%%%%%%
Equations (\ref{eq01}) are integrated in time using a 4th order Runge-Kutta 
algorithm with a constant time-step (typically $h=0.02$) with free-end boundary 
conditions, i.e., with 
\begin{equation}
\label{100}
   q_0(\tau) =0, \qquad q_{N+1}(\tau) =0,   
\end{equation}
that account for the termination of the structure in a finite system. In what 
follows, the system is initialized with a single-site excitation of amplitude 
$A$ at the interface, i.e., with 
\begin{equation}
\label{101}
   q_n(\tau=0) =A \, \delta_{n,n_e}, \qquad \dot{q}_n(\tau=0) =0,
\end{equation}
where $n_e =(N+1)/2$ and $\delta_{n,n_e}$ is the delta function ($n=1, ... ,N$).
%%%%%%%%%%%%%%%%%%%%%%%%%%%%%%%%%%%%%%%%%%%%%%%%%%%%%%%%%%%%%%%%%%%%%%%%%%%%%%%%
\begin{figure}[t!]
\includegraphics[angle=0,width=1.0 \linewidth]{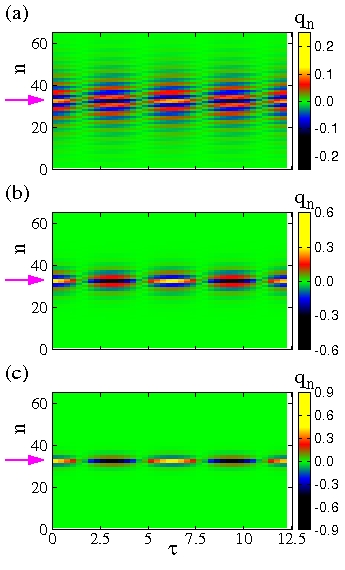}
\caption{(color online) 
Maps of the charge profiles $q_n$, where $n$ is the site number 
($n=1,2,3, ... ,N$), as a function of normalized temporal variable $\tau$ in the 
topologically protected localized interface state during two oscillation periods 
$2T =4\pi \simeq 12.56$, for $A=1$, $N=65$, $\lambda_M =-0.04$, $\gamma =0.005$, 
and 
(a) $\lambda_M' =-0.03$ ($\delta \lambda_M =0.01$); 
(b) $\lambda_M' =-0.02$ ($\delta \lambda_M =0.02$);  
(c) $\lambda_M' =-0.01$ ($\delta \lambda_M =0.03$). 
The (magenta) arrows indicate the position at $n =n_e =(N+1)/2$ of the central 
split-ring resonator (SRR) which has neither gain nor loss. 
}
\label{fig02}
\end{figure}
%%%%%%%%%%%%%%%%%%%%%%%%%%%%%%%%%%%%%%%%%%%%%%%%%%%%%%%%%%%%%%%%%%%%%%%%%%%%%%%%

In Figs. \ref{fig02}(a), (b), and (c), two-dimensional maps of the charge 
profiles $q_n$, where $n$ is the site number ($n=1,2,3, ... ,N$), are shown as a 
function of the normalized temporal variable $\tau$  during two oscillation 
periods $2 T$, where $T =2\pi$ ($\Omega =1$), for three dimerization strengths 
$\delta \lambda_M =0.03$, $0.02$, and $0.01$, respectively, and $\gamma =0.005$. 
In order to obtain these profiles, the topological PTMM is initialized according 
to Eq. (\ref{101}) with $A=1$, and then Eqs. (\ref{eq01}) are integrated in time 
for $\sim 10^5 ~T$ time-units (i.e., for $\sim 10^5$ periods), to eliminate 
transients and reach a steady state. Only the profiles obtained during the last 
two periods of integration are kept and mapped onto the $\tau-n$ plane in Fig. 
\ref{fig02}. These long transients originate from the fact that the initial 
condition is not an exact solution of the topological PTMM. Thus, some of its 
energy escapes away from the interface region during evolution and generates 
standing waves that may destroy localization. For obtaining more clear results, 
the SRRs with gain that belong to the dimers at the ends of the chain have been 
converted into lossy ones. That makes the system slightly lossy, but also helps 
the excessive energy to slowly dissipate and leave behind smooth profiles. These 
lossy ends are employed in all simulations of the dynamical equations Eqs. 
(\ref{eq01}) below.

As it is observed in Fig. \ref{fig02}, a localized state forms at the interface 
(whose position at $n=n_e$ is indicated by the arrows) and moreover its 
localization is stronger for higher $\delta \lambda_M$. This result is 
independent of the amplitude $A$, whose only effect is to decrease the maximum 
of the localized states proportionally to its magnitude, since the system is 
linear. In earlier works on 1D configurations similar to those of Fig. 
\ref{fig01} \cite{Schomerus2013,Poli2015,Blanco2016,Weimann2017}, it has been 
argued that by interfacing two dimer chains with different topologies (i.e., the 
topological and the non-topological one) and thus different topological 
invariants, results in a topological transition accompanied by a topologically 
protected state at the interface such as those presented here.    
%%%%%%%%%%%%%%%%%%%%%%%%%%%%%%%%%%%%%%%%%%%%%%%%%%%%%%%%%%%%%%%%%%%%%%%%%%%%%%%%
\begin{figure}[h!]
\includegraphics[angle=0,width=1.0 \linewidth]{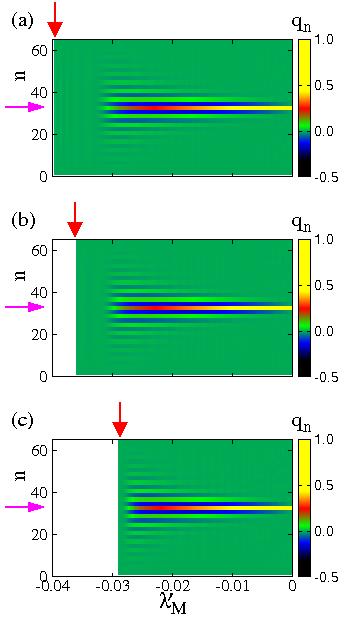}
\caption{(color online) 
Maps of the charge profiles $q_n$, where $n$ is the site number 
($n=1,2,3, ... ,N$), as a function of the inter-dimer coupling $\lambda_M'$ for 
$N=65$, $\lambda_M =-0.04$, and 
(a) $\gamma =0$; (b) $\gamma =0.005$; (c) $\gamma =0.01$.  
The horizontal (magenta) arrows indicate the position at $n =n_e =(N+1)/2$ of 
the central split-ring resonator (SRR) which has neither gain nor loss.
The vertical (red) arrows indicate the critical $\lambda_M'$ that separates the
exact from the broken $\cal PT$ phase, $(\lambda_M')_2$, whose values are 
$-0.04$, $-0.036$, and $-0.029$ for (a), (b), and (c), respectively. 
}
\label{fig03}
\end{figure}
%%%%%%%%%%%%%%%%%%%%%%%%%%%%%%%%%%%%%%%%%%%%%%%%%%%%%%%%%%%%%%%%%%%%%%%%%%%%%%%%

In Figs. \ref{fig03}(a), (b), and (c), the variation of the charge profiles 
$q_n$ of the topologically protected localized interface state is shown as a 
function of $\lambda_M'$ for $\gamma =0$, $0.005$, and $0.01$, respectively. 
In each subfigure, the topological PTMM is initialized according to Eq. 
(\ref{101}) with $A=1$ for each $\lambda_M'$, and then Eqs. (\ref{eq01}) are 
integrated in time for $\sim 10^5 ~T$ time-units for the transients to die out. 
The charge profiles at the end of the integration time for each $\lambda_M'$ are 
then mapped as a function of $\lambda_M'$. Note that lossy ends, as described 
above, are also employed here during time-integration.

For decreasing $\lambda_M'$ or equivalently decreasing $\delta \lambda_M$, the 
maximum of the profiles $q_n$ at $n =n_e$ (indicated by the horizontal magenta 
arrows) lowers smoothly approximatelly at the same rate in all subfigures. At a 
critical $\lambda_M' =(\lambda_M')_1$, however, the topologically protected 
states destabilize, leaving behind extended states which eventually vanish due 
to the lossy ends. With further decreasing $\lambda_M'$, in the case of 
$\gamma =0.005$ and $0.01$ (Figs. \ref{fig03}(b) and (c)), even these states 
destabilize at another critical $\lambda_M' =(\lambda_M')_2$ (indicated by the 
vertical red arrows), where a transition from the exact to the broken $\cal PT$ 
phase occurs. Obviously there is no such transition in the case $\gamma =0$ 
(Fig. \ref{fig03}(a)). The value of $(\lambda_M')_1$ is $-0.03275$, $-0.03175$, 
and $-0.0285$, while that of $(\lambda_M')_2$ is $-0.04$, $-0.036$, and $-0.029$, 
respectively, for $\gamma =0$, $0.005$, and $0.01$.

At $\lambda_M' =(\lambda_M')_1$, the topologically protected states break down 
due to dynamical instabilities arising independently of the existence or not of 
$\cal PT$ symmetry, at least for $\gamma$ relatively far from $\gamma_c$. The 
value of $(\lambda_M')_2$ generally depends on $\delta \lambda_M$, and it
decreases with decreasing $\gamma$. Thus, the instability at 
$\lambda_M' =(\lambda_M')_2$ observed in Figs. \ref{fig03}(b) and (c) signifies 
the transition of the topological PTMM to the broken $\cal PT$ phase where no 
stable states exist (the lossy ends cannot prevent this divergence). These 
results indicate that for the topological PTMM, the dimerization strength 
$\delta \lambda_M$ is the important parameter for the observation of protected 
topological localization. The gain-loss coefficient $\gamma$ only slightly 
affects the profiles $q_n$ of the protected states, as it can be inferred by 
comparing Fig. \ref{fig03}(a) with Figs. \ref{fig03}(b) and (c). Clearly, 
protected topological localization emerges even in the absence of $\cal PT$ 
symmetry, i.e., for $\gamma =0$ (Fig. \ref{fig03}(a)).

%%%%%%%%%%%%%%%%%%%%%%%%%%%%%%%%%%%%%%%%%%%%%%%%%%%%%%%%%%%%%%%%%%%%%%%%%%%%%%%%
\section{The Quadratic Eigenvalue Problem}
\begin{figure*}[t!]
\includegraphics[angle=0,width=1.0 \linewidth]{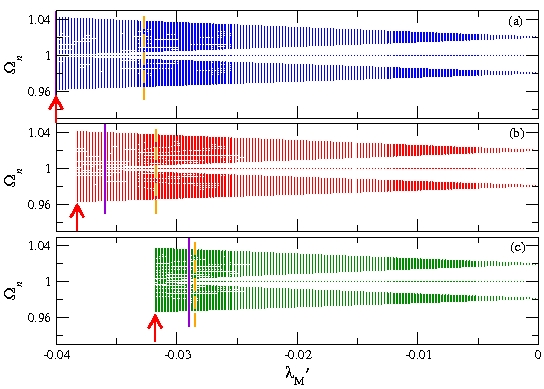}
\caption{(color online) 
Real eigenfrequencies $\Omega_n$ ($n=1,2,3, ... ,N$) as a function of the 
inter-dimer coupling $\lambda_M'$ for $N=65$, $\lambda_M =-0.04$, and 
(a) $\gamma =0$;  
(b) $\gamma =0.005$; 
(c) $\gamma =0.01$. 
The vertical (red) arrows indicate the critical $\lambda_M'$ that separate real 
from complex eigenvalues or equivalently the exact from the broken $\cal PT$ phase, 
$\widetilde{(\lambda_M')}_2$, whose values are 
$\widetilde{(\lambda_M')}_2 =-0.04$, $-0.03825$, and $-0.03175$ for (a), (b), 
and (c), respectively. The solid-maroon and dashed-orange vertical segments 
indicate the values of $(\lambda_M')_2$ and $(\lambda_M')_1$ discussed in the 
context of Fig. \ref{fig03}. 
}
\label{fig04}
\end{figure*}

\begin{figure}[h!]
\includegraphics[angle=0,width=1.0 \linewidth]{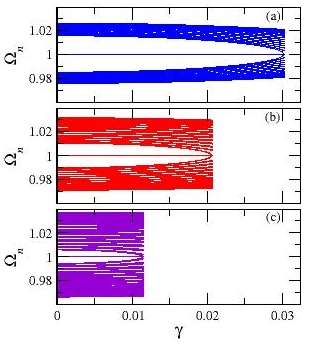}
\caption{(color online) 
Real eigenfrequencies $\Omega_n$ as a function of the gain-loss coefficient 
$\gamma$ for $N=65$, $\lambda_M =-0.04$, and
(a) $\lambda_M' =-0.01$ ($\delta \lambda_M =0.03$);
(b) $\lambda_M' =-0.02$ ($\delta \lambda_M =0.02$);
(c) $\lambda_M' =-0.03$ ($\delta \lambda_M =0.01$).
}
\label{fig05}
\end{figure}

%%%%%%%%%%%%%%%%%%%%%%%%%%%%%%%%%%%%%%%%%%%%%%%%%%%%%%%%%%%%%%%%%%%%%%%%%%%%%%%%
By substituting ${\bf q} = {\bf q}_0 \, e^{i \Omega \tau}$ into Eq. (\ref{eq01}) 
we get 
\begin{equation}
\label{eq11}
  \left\{-\Omega^2 \hat{\bf \Lambda}_M +i \Omega \hat{\bf \Gamma} 
  +\hat{\bf I}_{N\times N} \right\} {\bf q}_0 =\hat{\bf 0}, 
\end{equation}
where $\hat{\bf I}_{N\times N}$ is the $N\times N$ identity matrix. Eq. 
(\ref{eq11}) is a quadratic eigenvalue problem (QEP) that can be solved by 
standard eigenproblem solvers after its linearization by the classical 
augmentation procedure \cite{Duncan1935,Afolabi1987}. In the exact $\cal PT$ 
phase all the eigenfrequencies are real, and their spectrum consists of two 
bands separated by a gap, and an isolated mid-gap eigenfrequency equal to unity 
(Figs. \ref{fig04} and \ref{fig05}). The latter corresponds to a highly 
localized eigenmode.  
%%%%%%%%%%%%%%%%%%%%%%%%%%%%%%%%%%%%%%%%%%%%%%%%%%%%%%%%%%%%%%%%%%%%%%%%%%%%%%%%
%%%%%%%%%%%%%%%%%%%%%%%%%%%%%%%%%%%%%%%%%%%%%%%%%%%%%%%%%%%%%%%%%%%%%%%%%%%%%%%%

In Fig. \ref{fig04}, all the real eigenfrequencies are plotted as a function of 
$\lambda_M'$ for three values of $\gamma$. For relatively high $\lambda_M'$ as 
compared with $\lambda_M$ or equivalently relatively high $\delta \lambda_M$, 
two distinct frequency bands are formed. The bands are separated by a gap, in 
which a mid-gap frequency at $\Omega =1$ can be observed. However, for 
decreasing $\lambda_M'$ both bands widen until they merge together at a critical 
value $\lambda_M' =\widetilde{(\lambda_M')}_2$ whose actual value depends on 
$\gamma$. Below that point, the until then real eigenfrequencies acquire a 
nonzero imaginary part that signifies the transition from the exact to the 
broken $\cal PT$ phase where no stable solutions exist. In Fig. \ref{fig04}(a), 
for which $\gamma =0$, the two bands merge at $\lambda_M =\lambda_M'$ or 
equivalently at $\delta \lambda_M =0$ ($\widetilde{(\lambda_M')}_2 =-0.04$), 
while in Figs. \ref{fig04}(b) and (c) they merge at 
$\delta \lambda_M \simeq 0.00175$ ($\widetilde{(\lambda_M')}_2 \simeq -0.03825$) 
and $\simeq 0.00825$ ($\widetilde{(\lambda_M')}_2 \simeq -0.03175$), 
respectively. Note that in the presence of $\cal PT$ symmetry ($\gamma \neq 0$), 
the values of $\widetilde{(\lambda_M')}_2$ obtained by solving the QEP are 
smaller than the corresponding ones obtained by solving directly the dynamical 
equations Eqs. (\ref{eq01}), $(\lambda_M')_2$, with the same set of parameters. 
The reason for this difference is again dynamical instabilities peculiar to 
$\cal PT$ symmetric systems ($\widetilde{(\lambda_M')}_2 = (\lambda_M')_2$ for 
$\gamma =0$). In Fig. \ref{fig04}, the locations of $\widetilde{(\lambda_M')}_2$ 
are indicated by vertical red arrows, while the locations of $(\lambda_M')_1$ and 
$(\lambda_M')_2$ which were defined in the context of Fig. \ref{fig03} by 
dashed-orange and solid-maroon vertical segments, respectively.

In Fig. \ref{fig05}, the real eigenfrequencies of the topological PTMM  are 
plotted as a function of $\gamma$ for three values of $\delta \lambda_M$. The 
eigenfrequency spectrum is similar to that in Fig. \ref{fig04}. There are again 
two distinct frequency bands separated by a gap with a mid-gap eigenfrequency at 
$\Omega =1$. Here, the bandwidths do not change significantly for large 
intervals of $\gamma$ relatively far from $\gamma_c$ (e.g., from $\gamma =0$ to 
$\sim 0.015$ in Fig. \ref{fig05}(a)). The bandwidths are larger for lower 
$\lambda_M'$ (lower $\delta \lambda_M$), and the two bands merge together at 
$\gamma =\gamma_c$ whose value depends on $\lambda_M'$.

%%%%%%%%%%%%%%%%%%%%%%%%%%%%%%%%%%%%%%%%%%%%%%%%%%%%%%%%%%%%%%%%%%%%%%%%%%%%%%%%
\begin{figure}[h!]
\includegraphics[angle=0,width=0.95 \linewidth]{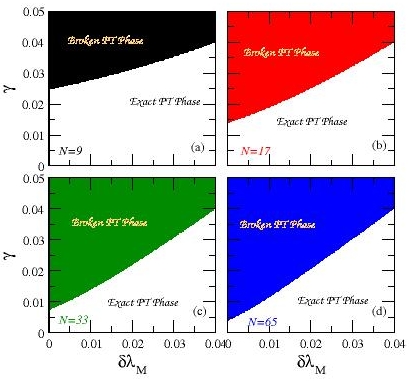}
\caption{(color online) 
$\cal PT$ phase diagrams for topological $\cal PT$ metamaterials with 
$\lambda_M =-0.04$ and (a) $N=9$; (b) $N=17$; (c) $N=33$; (d) $N=65$, on the 
$\delta \lambda_M - \gamma$ plane. Colored (white) areas indicate broken (exact) 
$\cal PT$ phase.
}
\label{fig06}
\end{figure}

%%%%%%%%%%%%%%%%%%%%%%%%%%%%%%%%%%%%%%%%%%%%%%%%%%%%%%%%%%%%%%%%%%%%%%%%%%%%%%%%
The boundary between the exact and the broken $\cal PT$ phase of the system is 
constructed numerically by solving the QEP. When all the eigenfrequencies are 
real, the PTMM is considered to be in the exact $\cal PT$ phase, otherwise it is 
considered to be in the broken $\cal PT$ phase. Such ``$\cal PT$ phase diagrams'' 
in the $\delta \lambda_M -\gamma$ plane are shown in Fig. \ref{fig06} for 
topological PTMMs with different $N$. These figures provide insights about the 
parameter values leading to the exact $\cal PT$ phase. For the values of $N$ 
used in Fig. \ref{fig06}, the parameter area providing exact $\cal PT$ phase and 
thus stability and topological effects shrinks with increasing $N$. For larger 
$N$, however, such $\cal PT$ phase diagrams do not change significantly with $N$ 
while the boundary between the two phases approaches the line 
$\gamma =\delta \lambda_M$. 
%%%%%%%%%%%%%%%%%%%%%%%%%%%%%%%%%%%%%%%%%%%%%%%%%%%%%%%%%%%%%%%%%%%%%%%%%%%%%%%%
%%%%%%%%%%%%%%%%%%%%%%%%%%%%%%%%%%%%%%%%%%%%%%%%%%%%%%%%%%%%%%%%%%%%%%%%%%%%%%%%
%%%%%%%%%%%%%%%%%%%%%%%%%%%%%%%%%%%%%%%%%%%%%%%%%%%%%%%%%%%%%%%%%%%%%%%%%%%%%%%%
\begin{figure}[h!]
\includegraphics[angle=0,width=1.0 \linewidth]{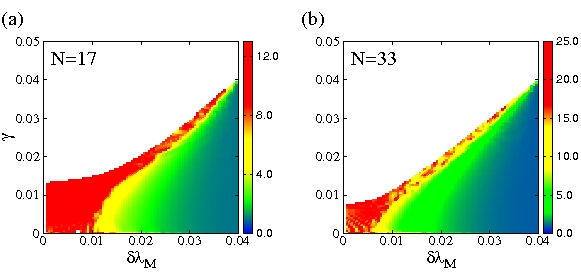}
\caption{(color online) 
The time-average of $P_e$ over a period $T=2\pi$, $\left< P_e \right>_{T=2\pi}$,
mapped on the $\delta \lambda_M - \gamma$ plane for $\lambda_M =-0.04$ and 
(a) $N=17$; (b) $N=33$. White areas indicate broken $\cal PT$ phase.
}
\label{fig07}
\end{figure}
%%%%%%%%%%%%%%%%%%%%%%%%%%%%%%%%%%%%%%%%%%%%%%%%%%%%%%%%%%%%%%%%%%%%%%%%%%%%%%%%
The results of Fig. \ref{fig06} can be compared to dynamic simulations for the 
{\em energetic participation number} 
\begin{equation}
\label{eq12}
   P_e =\frac{E_{tot}^2}{\sum_{n=1}^N E_n^2}
\end{equation}
of steady states of the system on the $\delta \lambda_M-\gamma$ plane, where 
\begin{eqnarray}
\label{eq13}
   E_n =\frac{1}{2} \left\{ \dot{q}_n^2 +{q}_n^2
~~~~~~~~~~~~~~~~~~~~~~~~~~~~~~~~~~~~~~~~~
\right. \nonumber \\ \left.
       +(\hat{\bf{\Lambda}}_M)_{n,n+1} \dot{q}_n \dot{q}_{n+1}
       +(\hat{\bf{\Lambda}}_M)_{n-1,n} \dot{q}_{n-1,n} \dot{q}_n 
        \right\}
\end{eqnarray}
and 
\begin{equation}
\label{eq14}
   E_{tot} =\frac{1}{2} \left\{ \dot{\bf q}^T \hat{\bf{\Lambda}}_M \dot{\bf q} 
           +{\bf q}^T \hat{\bf I}_{N\times N} {\bf q} \right\}
\end{equation}
are the approximate energy density and total energy, respectively, in which the 
terms proportional to $\gamma$ are neglected. In Fig. \ref{fig07}, the 
time-average of $P_e$ over $T=2\pi$ (i.e., for $\Omega =1$), 
$\left< P_e \right>_{T=2\pi}$, is mapped on the $\delta \lambda_M - \gamma$ plane 
for $N=17$ and $33$. At each point on that plane the system is initialized with 
$A=1$. The integration time is large enough to eliminate transients and lossy  
ends were employed. Figs. \ref{fig07}(a) and (b) provide a slightly lowered 
boundary between the exact and broken $\cal PT$ phase compared to that in Figs. 
\ref{fig06}(b) and (c), respectively. This is due to dynamical instabilities 
setting in when $\gamma$ approaches $\gamma_c$. Most importantly, 
$\left< P_e \right>_{T=2\pi}$ quantifies the degree of localization of the 
topologically protected states, since it roughly measures the number of the 
energetically strongest excited SRRs. In a sense, it quantifies the degree of 
topological protection. The highest degree of localization 
($\left< P_e \right>_{T=2\pi} \sim 1-2$) is obtained in the blue areas, where 
all the energy of a state in this part of the plane is concentrated in one or 
two SRRs only. The degree of localization then gradually decreases in the green 
and yellow areas, and practically vanishes in the red areas (in the white areas, 
stable solutions do not exist). Note that for fixed $\delta \lambda_M$, the 
degree of localization decreases with increasing $\gamma$.

%%%%%%%%%%%%%%%%%%%%%%%%%%%%%%%%%%%%%%%%%%%%%%%%%%%%%%%%%%%%%%%%%%%%%%%%%%%%%%%%
\section{Conclusions}
%%%%%%%%%%%%%%%%%%%%%%%%%%%%%%%%%%%%%%%%%%%%%%%%%%%%%%%%%%%%%%%%%%%%%%%%%%%%%%%%
A topological PTMM having the configuration in Fig. \ref{fig01} is considered, 
using equivalent circuit modeling in which the dynamics is described by a set of 
linear second order differential equations for the charges $q_n$. Direct 
numerical simulations demonstrate the existence of interface topologically 
protected localized states in the exact $\cal PT$ phase, that oscillate with 
the mid-gap frequency $\Omega =1$ of the eigenfrequency spectum. The results 
presented here indicate that the {\em dimerization strength} $\delta \lambda_M$ 
is the important parameter regarding topological protection of interface 
localized states. Conversely, the effect of $\gamma$ is much weaker as long as 
its value is relatively far from the critical one, $\gamma_c$; its presence is 
not necessary for topological protection and even disfavors slightly 
localization. However, $\cal PT$ symmetry does not destroy the topological 
nature of the localized interface state and thus such $\cal PT-$symmetric states
can be realized. Given that robust topological edge states have been already 
observed in 1D SRR chains \cite{JJiang2018}, the predictions above could be 
experimentally confirmed at least for $\gamma=0$.

%%%%%%%%%%%%%%%%%%%%%%%%%%%%%%%%%%%%%%%%%%%%%%%%%%%%%%%%%%%%%%%%%%%%%%%%%%%%%%%%
\section*{Acknowledgements}
The authors gratefully acknowledge the financial support of the Ministry of 
Science and Higher Education of the Russian Federation in the framework of 
Increase Competitiveness Program of NUST ``MISiS'' (No. K2-2019-010), 
implemented by a governmental decree dated 16th of March 2013, N 211, and also 
thank the Superconducting Metamaterials Laboratory for its hospitality during 
visits. NL acknowledges support by the General Secretariat for Research and 
Technology (GSRT) and the Hellenic Foundation for Research and Innovation (HFRI) 
(Grant No. 203).
%%%%%%%%%%%%%%%%%%%%%%%%%%%%%%%%%%%%%%%%%%%%%%%%%%%%%%%%%%%%%%%%%%%%%%%%%%%%%%%%

%%%%%%%%%%%%%%%%%%%%%%%%%%%%%%%%%%%%%%%%%%%%%%%%%%%%%%%%%%%%%%%%%%%%%%%%%%%%%%%%
%\bibliography{BibTex-Library-11Nov2019.bib}
%\bibliographystyle{unsrt}
%%%%%%%%%%%%%%%%%%%%%%%%%%%%%%%%%%%%%%%%%%%%%%%%%%%%%%%%%%%%%%%%%%%%%%%%%%%%%%%%

%%%%%%%%%%%%%%%%%%%%%%%%%%%%%%%%%%%%%%%%%%%%%%%%%%%%%%%%%%%%%%%%%%%%%%%%%%%%%%%%
\end{document}